\documentclass[aps,pre,showpacs,amsmath,twocolumn]{revtex4}

\usepackage{graphicx}
\usepackage{dcolumn}
\usepackage{bm}
\usepackage{color}

\usepackage{amssymb}

\newcommand\beq{\begin{equation}}
\newcommand\eeq{\end{equation}}
\newcommand\beqa{\begin{eqnarray}}
\newcommand\eeqa{\end{eqnarray}}
\newcommand{\nn}{\nonumber\\}

\newcommand{\HS}{\text{HS}}
\newcommand{\PS}{\text{PS}}

\begin{document}
\title{Molecular dynamics simulation study of self-diffusion for penetrable-sphere model fluids}
\author{Soong-Hyuck Suh}
\email{shsuh@kmu.ac.kr}
\author{Chun-Ho Kim}
\affiliation{Department of Chemical Engineering, Keimyung University, Daegu 704-701, Korea}
\author{Soon-Chul Kim}
\email{sckim@andong.ac.kr}
\affiliation{Department of Physics, Andong National University, Andong 305-600, Korea}
\author{Andr\'es Santos}
\email{andres@unex.es}
\homepage{http://www.unex.es/eweb/fisteor/andres/}
\affiliation{Departamento de F\'{\i}sica, Universidad de Extremadura, E-06071
Badajoz, Spain}

\begin{abstract}
Molecular dynamics simulations are carried out to investigate the diffusion behavior of penetrable-sphere model fluids characterized by a finite energy barrier $\epsilon$. The self-diffusion coefficient is evaluated from the time-dependent  velocity autocorrelation function and  mean-square displacement. Detailed insights into the cluster formation for penetrable spheres are gained from the Enskog factor, the effective particle volume fraction, the mean free path, and the collision frequency for both the soft-type penetrable and the hard-type reflective collisions. The  simulation data are compared to theoretical predictions from the Boltzmann kinetic equation and from a simple extension to finite $\epsilon$ of the Enskog prediction for impenetrable hard spheres ($\epsilon\to\infty$). A reasonable agreement between theoretical and simulation results is found in the cases of  $\epsilon^*\equiv \epsilon/k_BT=0.2$, $0.5$, and $1.0$. However, for dense systems (packing fraction $\phi>0.6$) with a highly repulsive energy barrier ($\epsilon^*=3.0$), a poorer agreement was observed due to metastable  static effects of clustering formation and  dynamic effects of correlated collision processes among these cluster-forming particles.
\end{abstract}


\date{\today}
\maketitle

\section{Introduction\label{sec1}}

The so-called penetrable-sphere (PS) pair potential is defined as
\beq
u^\PS(r)=\begin{cases}
\epsilon,& r<\sigma,\\
0,&r>\sigma,
\end{cases}
\label{1.1}
\eeq
where $\sigma$  is the diameter of the penetrable spheres and $\epsilon>0$ is  the strength of the repulsive energy barrier between two overlapping spheres when they penetrate each other.
The PS model was suggested by Marquest and Witten \cite{MW89} as a simple theoretical approach to micelles in a solvent to explain the experimentally observed crystallization of copolymer mesophases, where a simple cubic solid phase coexisted with the disordered suspension. This simple  model system has been the subject of several theoretical and simulation studies \cite{LWL98,S99,FLL00,SF02,CG03,AS04,MS06,SM07,MYS07}. An excellent review in this area up to 2001 can  be found  in Ref.\ \cite{L01}. More recently, the PS model has been extended to include a short-range attractive tail \cite{SFG08,FGMS09,FGMS10}.

The PS  interaction model  reduces to the classical hard-sphere (HS) system when  $\epsilon^*\equiv \epsilon/k_BT\to\infty$ (where $T$ is the temperature and $k_B$ is the Boltzmann constant). This is equivalent to the zero-temperature limit $T^*\equiv k_BT/\epsilon\to 0$. In the opposite high-penetrability or high-temperature limit ($\epsilon^*\to 0$, $T^*\to\infty$) the PS system  becomes a collisionless ideal gas. Except in the pure HS case, penetrability allows one in principle to consider systems with any value of the nominal packing fraction $\phi \equiv (\pi/6)n\sigma^3$, where $n$ is the number density. The first correction to the equilibrium ideal-gas structural and thermodynamic properties in the combined limit $\epsilon^*\to 0$, $\phi\to \infty$ with $\epsilon^*\phi=\text{const}$ has been exactly obtained \cite{AS04}.

Recently, the nonequilibrium transport properties of unbounded potentials, such as the linear-core \cite{DAG09} and Gaussian-core \cite{SE10} fluids have received much attention. In the case of the PS model, the Liouville operator and the Boltzmann–-Lorentz
collision operator  were long time ago derived in Refs.\ \cite{A75} and \cite{GDL79}, respectively. As a more explicit application of kinetic theory to soft matter \cite{S05}, one of the authors  has derived the relevant transport coefficients (self-diffusion, shear viscosity, and thermal conductivity) in the context of the Chapman--Enskog method for the Boltzmann equation of dilute gases ($\phi\ll 1$) \cite{CC70}. Interestingly, in the PS binary collision dynamics the particle penetration is analogous to the double refraction of light through a sphere made of a transparent material of relative refraction index depending on the relative collision velocity and the repulsive energy parameter \cite{S05}.

In addition to transport properties, two of us \cite{KS02} have applied two different theoretical predictions, based on the fundamental-measure theory proposed by Schmidt \cite{S99} and the bridge density-functional approximation proposed by Zhou and Ruckenstein \cite{ZR00}, to the inhomogeneous structure of PS model fluids in the spherical pore system. More recently \cite{KSS09}, as a continuation of theoretical approaches along this direction, the modified density-functional theory (based on both the bridge density functional and the contact-value theorem) has been investigated for the structural properties of PS fluids near a slit hard wall and the Verlet-modified bridge function for one-component systems proposed by Choudhury and Ghosh \cite{CG03} has also been extended to PS fluid mixtures.

Theoretical results for such a precisely defined model potential can be directly compared against the \emph{exact} machine-experimental data obtained from computer simulations via Monte Carlo (MC) and molecular dynamics (MD) methods. So far, almost all simulations for the PS model fluid have been carried out using the MC method. On the other hand,  better statistics can be achieved in MD simulations, particularly in systems with discontinuous interactions. For instance, in order to calculate the virial route to the equation of state for hard-core systems, MC computations require an accurate estimation of the radial distribution function at the contact point. Computationally, the pair distribution function may change rapidly near the contact distance in the systems of ionic solutions, highly charged colloids, aligned liquid crystals, etc. Under those conditions the extrapolation of the pair distribution function to the contact value may lead to large uncertainties. For this reason, the pressure  determined by MC simulations is known to be less accurate than that by MD computations \cite{AT87}.

The main motivation in the present work is to undertake a detailed molecular-based simulation study of transport properties of PS model fluids, which can be in turn directly  compared with  theoretical and/or  empirical predictions. MD data for the PS interaction potential have not been presented in the literature before. More specifically,  we have investigated the PS model system via the equilibrium MD method over a wide range of densities and repulsive energy parameters to investigate the equation of state, the self-diffusion coefficient, and its related time-dependent quantities including the velocity autocorrelation function (VACF) and the mean-square displacement (MSD). Together with existing approximations for the equilibrium equation of state as well as for the self-diffusion coefficient in the Boltzmann and Enskog-like  descriptions, our simulation results can be used to assess and construct {a} fundamental basis of theoretical and practical predictions for the relevant transport properties. Such simulation approaches at the atomic or molecular level can  also be used in improved statistical integral-equation theories of liquid state and help in interpreting real experimental data.

The organization of this paper is as follows.
Section \ref{sec2} presents some simple theoretical approximations for the equation of state and the self-diffusion coefficient of the PS fluid. The MD method employed in this paper is briefly described in Sec.\ \ref{sec3}. The most important part of the paper is contained in Sec.\ \ref{sec4}, where the simulation results for the pressure, the self-diffusion coefficient, the Enskog $\chi$ factor,  the effective particle volume fraction, the mean free path, the collision frequencies, the velocity autocorrelation function, and the mean-square displacement are presented, compared with theoretical approaches, and discussed. The paper is closed with some concluding remarks in Sec.\ \ref{sec5}.

\section{Theoretical approaches for the PS fluid\label{sec2}}

\subsection{Equation of state\label{subsec2a}}

The equilibrium compressibility factor $Z=p/nk_BT$ in the PS model, where $p$ is the pressure, is given by \cite{SM07}
\beq
Z^\PS=1+4\phi x \chi^\PS,
\label{2.1}
\eeq
where
\beq
x\equiv 1-e^{-\epsilon^*}
\label{x}
\eeq
is a parameter measuring the degree of penetrability of the particles (ranging from $x=0$ in the free-penetrability limit $\epsilon^*\to 0$ to $x=1$ in the opposite impenetrability limit $\epsilon^*\to\infty$) and $\chi^\PS\equiv g^\PS(\sigma^+)$ is the contact value of the radial distribution function $g^\PS(r)$.

In Ref.\ \cite{MYS07} two approximate theories were proposed to obtain $g^\PS(r)$: one valid in the high-penetrability (i.e., small-$\epsilon^*$) regime and the other one in the low-penetrability (i.e., large-$\epsilon^*$) regime. We will refer to these two theories as the high-penetrability approximation (HPA) and the low-penetrability approximation (LPA), respectively. We give below the expressions for $\chi$ in both approximations.

In the HPA, $\chi$ is given by \cite{MYS07}
\beq
\chi^\PS=1+x w(1) e^{x w(1)},
\label{2.2}
\eeq
where
\beq
w(r)=\frac{48\phi x}{\pi r}\int_0^\infty dk\, \frac{\left(k\cos k-\sin k\right)^2}{k^3-24\phi x \left(k\cos k-\sin k\right)}\frac{\sin (kr)}{k^2}.
\label{2.3}
\eeq
Equation \eqref{2.2} reduces to the exact result $\chi=1+x w(1)$ in the limit $x\to 0$ (i.e., $\epsilon^*\to 0$) with  $\phi x=\text{const}$.

In the case of the LPA, one has \cite{MYS07}
\beq
\chi^\PS=\frac{A}{x}\frac{1+\phi/2}{(1-\phi)^2+(1-A)\phi(4-\phi)}.
\label{2.4}
\eeq
Here, $A$ is obtained from the transcendental equation
\beqa
12\frac{(1-x)A^2}{x(1-A)}\frac{\phi(1+\phi/2)}{(1-\phi)^2+(1-A)\phi(4-\phi)}\nn
\quad=
\sum_{i=1}^3\frac{z_ie^{z_i}}{B_1+2B_2z_i+3B_3z_i^2},
\label{2.4b}
\eeqa
where
$z_i$ ($i=1,2,3$)  are the three roots of the cubic equation
$1-B_1 z-B_2 z^2-B_3 z^3=0$ and the coefficients $B_1$, $B_2$, and
 $B_3$ are
 \begin{subequations}
 \label{2.5}
\beq
B_1=\frac{3}{2}\frac{\phi}{1+2\phi},\quad B_2=\frac{1}{2}\frac{1-\phi}{1+2\phi},
\label{2.5a}
\eeq
\beq
 B_3=\frac{1}{12}\left(\frac{1}{\phi A}-\frac{4-\phi}{1+2\phi}\right).
\label{2.5b}
\eeq
\end{subequations}
In the limit $x\to 1$ (i.e., $\epsilon^*\to \infty$), the solution of Eq.\ \eqref{2.4b} is $A=1$, and so one recovers the solution of the Percus--Yevick integral equation for HS \cite{HM06}.

\subsection{Self-diffusion coefficient\label{subsec2b}}
\begin{figure}
\includegraphics[width=0.7\columnwidth]{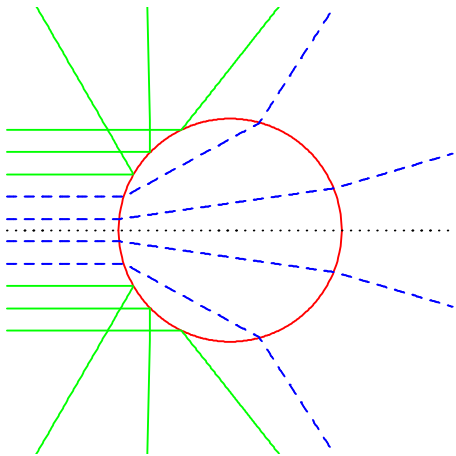}
\caption{(Color online) Different possible trajectories in a collision with a scaled relative velocity $v_{12}^*=1.1$. The dashed trajectories correspond to ``soft'' (refraction) collisions, while the solid {trajectories} correspond to ``hard'' (specular) collisions. In general, the collisions are of hard type if $v_{12}^*\sqrt{1-(b/\sigma)^2}<1$. As the (reduced) repulsive energy barrier $\epsilon^*$ increases, less and less  collisions are soft.\label{sketch}}
\end{figure}

In the low-density regime ($\phi\to 0$) the transport coefficients of a gas made of particles interacting via a potential $u(r)$ can be derived by application of the Chapman--Enskog method to the well-known Boltzmann equation \cite{CC70}. The crucial quantity distinguishing an interaction potential from another one is the scattering angle as a function of both the impact parameter $b$ and the relative velocity of the colliding pair, $v_{12}$.

In the particular case of the PS potential \cite{S05}, if the scaled relative velocity $v_{12}^*\equiv v_{12} /\sqrt{\epsilon/m}$, where $m$ is the mass of a particle, is smaller than 1, the ``projectile'' particle does not have enough kinetic energy to penetrate into the core of the ``target'' particle, and consequently it is deflected exactly as if the target were a hard sphere. On the other hand, if $v_{12}^*> 1$ and $b/\sigma<\sqrt{1-{v_{12}^*}^{-2}}$, i.e., if $v_{12}^*\sqrt{1-(b/\sigma)^2}>1$, the projectile traverses the core analogously to the double refraction of light through a sphere made of a transparent material of relative refraction index  $\sqrt{1-{v_{12}^*}^{-2}}$. In fact, if $b/\sigma>\sqrt{1-{v_{12}^*}^{-2}}$, i.e., $1<v_{12}^*<1/\sqrt{1-(b/\sigma)^2}$, a ``total reflection'' process takes place and the deflection is again as in the HS case.
Figure \ref{sketch} shows different possible trajectories corresponding to $v_{12}^*=1.1$.

In the first Sonine approximation, the self-diffusion coefficient $D_0$ obtained from the Boltzmann equation for the PS model is given by \cite{S05}
\beq
D_0^\PS=\frac{1}{16}\sqrt{\frac{\pi k_BT}{m}}\frac{\sigma}{\phi}\frac{1}{\Omega_{11}^*},
\label{2.6}
\eeq
where
\beq
\Omega_{11}^*
=1-\int_{\epsilon^*}^\infty dy\,
e^{-y}y^2R_{1}\left(y/\epsilon^*\right),
\label{2.7}
\eeq
with
\beqa
R_1(y)&=&\frac{\left({y}-1\right)\left({y}+2\right)}{6{y}^2}+\frac{4{y}^2-4{y}+3}{12{y}^{3/2}\sqrt{{y}-1}}\nn
&&
+\frac{2{y}-1}{8{y}^2\left({y}-1\right)}\ln\left[2{y}-2\sqrt{{y}({y}-1)}-1\right].
\label{2.8}
\eeqa

Obviously, in the low-penetrability limit  ($\epsilon^*\to\infty$), the self-diffusion coefficient for the PS model in Eq.\ \eqref{2.6} reduces to that of the HS model, namely
\beq
D_0^\HS=\frac{1}{16}\sqrt{\frac{\pi k_BT}{m}}\frac{\sigma}{\phi}.
\label{2.9}
\eeq

As said above, Eqs.\ \eqref{2.6} and \eqref{2.9} are derived from the Boltzmann equation (in the first Sonine approximation), and thus they are well justified in the high-dilution limit $\phi\to 0$. On the other hand, they do not account for finite-density effects. To correct this deficiency, several empirical or semi-empirical expressions have been proposed in the case of the HS model. Among them, the most basic one is provided by the Enskog kinetic theory \cite{CC70}. The Enskog correction for the self-diffusion coefficient in the HS system is
\beq
D^\HS=\frac{D_0^\HS}{\chi^\HS},
\label{2.10}
\eeq
where the Enskog  factor $\chi^\HS$  is the contact value of the radial distribution function of the HS fluid. This quantity is related to the corresponding equation of state in terms of the compressibility factor by
\beq
Z^\HS=1+4\phi\chi^\HS.
\label{2.11}
\eeq
Equation \eqref{2.10} takes into account that the effective number of collisions in a dense gas is increased by a factor $\chi^\HS$. Consequently, the self-diffusion coefficient is decreased by the same factor, relative to the Boltzmann prediction at the same density.
An excellent approximation for $\chi^\HS$ within the stable fluid region ($0\leq \phi\leq 0.494$) is provided by the Carnahan--Starling (CS) formula \cite{HM06,CS69}
\beq
\chi^\HS=\frac{1-\phi/2}{(1-\phi)^3}.
\label{CS}
\eeq

There are also a number of empirical formulas for the $D^\HS$. For  systems of $500$ HS particles or slightly fewer, the following analytical fit to MD data was reported by Speedy \cite{S87}:
\beq
D^\HS=D_0^\HS\left(1-\frac{\phi}{\phi_g}\right)\left[1+c_1 \left(\frac{\phi}{\phi_g}\right)^2-c_2 \left(\frac{\phi}{\phi_g}\right)^4\right].
\label{2.12}
\eeq
Here, $\phi_g\simeq 0.57$ is the packing fraction at the HS glass transition and Speedy's values are $c_1= 0.48$ and $c_2=1.17$.
Recently, a much more extensive MD computation was performed by  Sigurgeirsson and Heyes \cite{SH03} with an efficient MD algorithm dealing with up to $32000$ HS particles. They refined the values of the fitting coefficients $c_1$ and $c_2$ in Eq.\ \eqref{2.12} as $c_1=0.4740$ and $c_2=1.1657$.
The empirical form \eqref{2.12} takes into account the crowding effects in the first bracket term and the hydrodynamic back-flow effects at intermediate densities in the second bracket term. Both the Enskog formula \eqref{2.10} and the empirical expression \eqref{2.12} have in common the fact that, as expected on physical grounds, $D^\HS<D_0^\HS$, i.e., the self-diffusion coefficient decays more rapidly than hyperbolically with increasing density.

In the case of the PS system, the task of extending the Boltzmann result \eqref{2.6} to finite density to get the self-diffusion coefficient $D^\PS$ is much more difficult than in the HS case. In fact, the ratio $D^\PS/D_0^\PS$ is not only a function of density (as in the HS case) but also a function of temperature or, equivalently, of $\epsilon^*$. Based on the Enskog result  \eqref{2.10}, it seems natural to propose the following Enskog-like expression:
\beq
D^\PS=\frac{D_0^\PS}{\chi^\PS},
\label{2.13}
\eeq
where $D_0^\PS$ is given by Eqs.\ \eqref{2.6}--\eqref{2.8} and $\chi^\PS$ is either obtained from the empirical values of $Z^\PS$ from Eq.\ \eqref{2.1} or derived from the HPA [Eqs.\ \eqref{2.2} and \eqref{2.3}] or from the LPA [Eqs.\ \eqref{2.4}--\eqref{2.5}]. According to Eq.\ \eqref{2.13}, $D^\PS<D_0^\PS$. However, as will be seen in Sec.\ \ref{sec4}, this inequality is in general not supported by our MD data.

\section{Computational method \label{sec3}}

As a successful diagnostics tool, molecular-based computer simulations are usually employed to investigate the underlying diffusion behavior of the model system of interest. To this end, in this work we have carried out microcanonical MD simulations for the PS model fluid in a manner similar to that originally proposed by Alder and Wainwright for hard-core systems \cite{AW59}, which is well described elsewhere \cite{AT87}. Post-collision velocities for colliding pairs of particles are assigned according to the type of collision (see Fig.\ \ref{sketch}): either  hard-type specular reflection or  soft-type refraction. In all cases both the total  momentum and kinetic energy  are conserved in these PS collision conditions.

The initial configurations with 864 penetrable spheres were generated by randomly inserting particles with velocities drawn from Maxwell--Boltzmann  distributions. The initial configurations were aged, or equilibrated, for $5\times 10^7$ collisions before accumulating the final simulation results. Additional ensemble averages were evaluated from a total number of $5\times 10^8$ collisions.

Our MD algorithm has been tested in a number of ways. When the repulsive energy parameter was assigned  a large value (typically $\epsilon^*>3$) at the low-density regime ($\phi < 0.2$), the static and dynamic results generated from our MD simulations faithfully reproduced the pure HS system. Our resulting MD calculations for a few selected runs were also compared with MC computations reported in the literature. A good agreement with previous MC data for the thermodynamic and structural properties \cite{MYS07} again confirmed the validity of the MD method employed in this work. All MD results reported here are scaled to dimensionless quantities by using a unit particle diameter $\sigma$, a unit particle mass $m$, and a unit thermal energy $k_BT$. In these system units the reduced self-diffusion coefficient is expressed as $D^*=D/\sigma\sqrt{k_BT/m}$.

\section{Results and discussion \label{sec4}}
\begin{figure}
\includegraphics[width=0.950 \columnwidth]{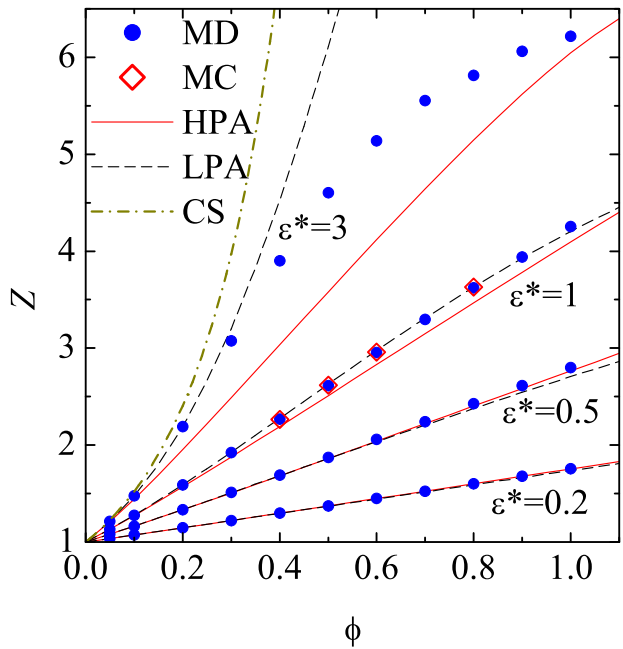}
\caption{(Color online) Compressibility factor $Z$ versus the packing fraction $\phi$ for several values of $\epsilon^*$. The circles are MD results, the solid lines are the HPA predictions [Eqs.\ \eqref{2.1}--\eqref{2.3}] and the dashed lines are the LPA predictions [Eqs.\ \eqref{2.1} and \eqref{2.4}--\eqref{2.5}]. The diamonds at $\epsilon^*=1$ and $\phi=0.4$, $0.5$, $0.6$, and $0.8$ are MC simulation data from Ref.\ \protect\cite{MYS07}. For comparison, the curve corresponding to the HS fluid described by the CS equation of state [Eqs.\ \protect\eqref{2.11} and \protect\eqref{CS}] is also plotted (dash-dotted line).
\label{Z}}
\end{figure}
\subsection{Equation of state}
Before presenting the results for the self-diffusion coefficient, which is the main quantity of interest in this work, it is worth considering the equation of state.

Figure \ref{Z} compares the MD simulation data for $Z$ with the HPA and the LPA. For the most penetrable case ($\epsilon^*=0.2$) both theories agree well with the MD data, with the agreement being excellent in the case of the HPA (relative deviations between MD and HPA smaller than $0.1\%$). For $\epsilon^*=0.5$ the HPA is still quite good, while for $\epsilon^*=1$ the LPA performs very well.
It is remarkable that both theories, while being based on opposite approaches, are so close each other up to $\epsilon^*\approx 1$ and densities as large as $\phi=1$.
In the least penetrable case ($\epsilon^*=3$) the LPA behaves reasonably well up to $\phi=0.3$ (where the PS system is only slightly distinguishable from a HS system, {represented here by the CS equation of state}) but strongly overestimates the MD values for larger densities, when penetrability effects start to play a relevant role. By accident the HPA does a good job for $\epsilon^*=3$ and $\phi\approx 1$.
It is important to remark that the MD data for the PS fluid with $\epsilon^*=3$ clearly deviate from the HS values for $\phi>0.2$. The reason is that, even though the value $\epsilon^*=3$ represents a rather high energy barrier, as the density increases more particles are forced to overlap, which results in a substantial decrease in  the pressure relative to that of the  HS system {at the same density}.

\subsection{Self-diffusion coefficient}
\begin{figure}
\includegraphics[width=0.95\columnwidth]{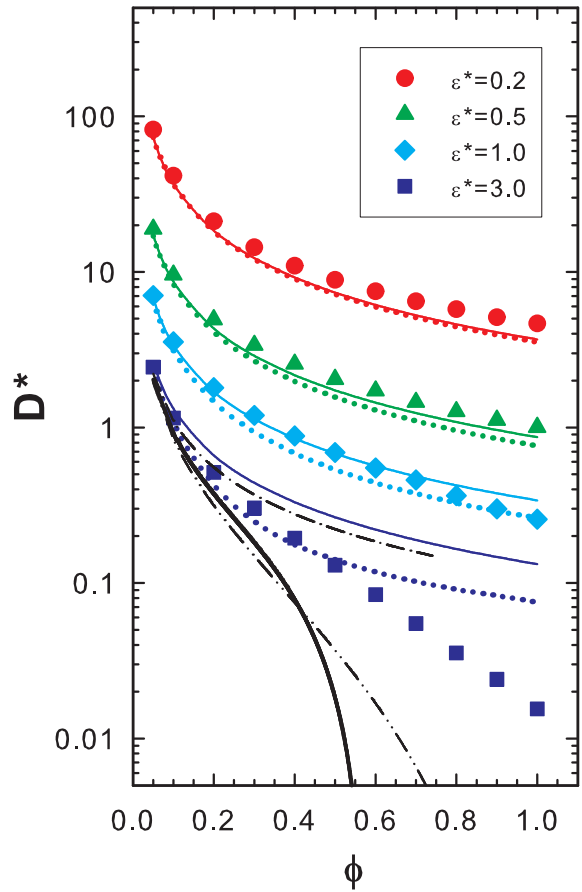}
\caption{(Color online) Reduced self-diffusion coefficient $D^*$ as a function of the packing fraction $\phi$ for several values of $\epsilon^*$. The symbols are MD simulations for the PS system, the solid lines are the Boltzmann theoretical predictions in the high-dilution limit for the PS fluid [cf.\ Eqs.\ \eqref{2.6}--\eqref{2.8}]; the dotted lines are the Enskog-like theoretical approximations for the PS fluid [cf.\ Eq.\ \eqref{2.13}],  complemented with the empirical values of $\chi^\PS$; the dash-dotted line {(interrupted at the maximum packing fraction of HS systems, $\phi=\pi\sqrt{2}/6\simeq 0.7405$)} represents the Boltzmann theoretical values in the high-dilution limit for the HS fluid [cf.\ Eq.\ \eqref{2.9}]; the dash-dot-dotted line represents the Enskog theoretical values for the HS fluid [cf.\ Eq. \eqref{2.10}], complemented with the CS values of $\chi^\HS$; and the thick solid line represents the empirical fitting-values from  MD simulations for the HS system [cf.\ Eq.\ \eqref{2.12} with  $c_1=0.4740$ and $c_2=1.1657$].
\label{fig1}}
\end{figure}

The self-diffusion coefficient is a single-particle quantity which has been more frequently studied in MD calculations than other collective transport properties, such as the shear viscosity and the thermal conductivity. The self-diffusion coefficient {$D$} can be determined from the temporal integration of the VACF using the Green--Kubo formula
\beq
D=\frac{1}{3}\int_0^\infty dt\, \langle\mathbf{v}(0)\cdot\mathbf{v}(t)\rangle
\label{VACF}
\eeq
and also from the slope of the MSD versus time using the Einstein relation
\beq
\langle r^2(t)\rangle =6Dt.
\label{MSD}
\eeq
In our MD simulations these two methods have produced consistent results, typically with less than 3\% differences.

By using a semi-logarithmic scale, in Fig.\ \ref{fig1} we  display the reduced self-diffusion coefficient $D^*$ as a function of the packing fraction $\phi$, as obtained from MD simulations, from the Boltzmann theoretical predictions in the high-dilution limit [Eqs.\ \eqref{2.6}--\eqref{2.8}] and from the Enskog-like theoretical approximations [Eq.\ \eqref{2.13}]. For comparison, Fig.\ \ref{fig1} also includes curves corresponding to the HS system ($\epsilon^*\to\infty$): the Boltzmann theoretical values in the high-dilution limit  [Eq.\ \eqref{2.9}], the Enskog theoretical values [Eq. \eqref{2.10}], and the empirical fitting-values from  MD simulations [Eq.\ \eqref{2.12}] with  $c_1=0.4740$ and $c_2=1.1657$.

There are several interesting features that can be observed in Fig.\ \ref{fig1}. As expected, the self-diffusion coefficient decreases with increasing density. It also decreases with increasing repulsive energy barrier at fixed $\phi$. This behavior is not counterintuitive. As the (reduced)  energy barrier $\epsilon^*$ increases, more and more collisions are of hard (specular) type. This gives rise to an increase in the effective collision frequency and thus a decrease in diffusion. At the Boltzmann level of description, this effect can be grasped from comparison between Eqs.\ \eqref{2.6} and \eqref{2.9}  and the property $\Omega_{11}^*<1$. For given values of $\epsilon^*$ and $\phi$, the self-diffusion coefficient evaluated from the Boltzmann kinetic equation is larger than the one from the Enskog-type correction equation. As shown by Eq.\ \eqref{2.13}, both coefficients are simply related by the $\chi$ factor, which is the contact value of the radial distribution function, i.e., $\chi=g(\sigma^+)$. This  value is  close to $1.0$ in the high-dilution limit and  increases with increasing density due to particle crowding effects near the contact distance.

Except for the case $\epsilon^*=3.0$, the qualitative behavior of the MD self-diffusion data is similar for the different values of $\epsilon^*$. In the cases   $\epsilon^*=0.2$, $0.5$, and $1.0$, we observe in Fig.\ \ref{fig1} that the PS Boltzmann  approximation  produces a reasonably good agreement with the corresponding MD data, even for relatively large values of the packing fraction (where the Boltzmann approximation would be expected to break down). In the case $\epsilon^*=0.2$ the relative deviations of the Boltzmann predictions with respect to the MD data  increase with density. For $\epsilon^*=0.5$, however, the relative deviations reach a maximum at about $\phi=0.6$ and then slowly decay with increasing density. In the two previous cases the Boltzmann prediction [Eqs.\ \eqref{2.6}--\eqref{2.8}] \emph{underestimates} the self-diffusion coefficient, at least in the density domain $0<\phi\leq 1$. This qualitative feature changes in the case $\epsilon^*=1.0$, where the Boltzmann values are below the MD ones up to $\phi\simeq 0.5$ only. In fact, the best general agreement between the Boltzmann results and the simulation data occurs, {because of this accidental crossing,}  for $\epsilon^*=1.0$. {It is expected that, on increasing the barrier, the crossing shifts toward lower densities.}
For the highly repulsive system with  $\epsilon^*=3.0$, {there is no crossing effect, and so} the  Boltzmann approximation \emph{overestimates} the MD values of $D^*$, {being} reliable only in the narrow range of densities $\phi\lesssim 0.2$.

For the pure HS fluid, it has been reported \cite{SH03,AGW70} that reliable self-diffusion data, significantly improving over the Boltzmann values, are obtained from the Enskog kinetic equation within the range of equilibrium stable HS fluids ($\phi <0.494$). One may see this point from Fig.\ \ref{fig1} by comparing the MD-fitting diffusion data with the HS Boltzmann and Enskog theoretical approximations.
It would then seem natural to expect a similar improvement of the Enskog-like correction \eqref{2.13} over the bare Boltzmann prediction also in the case of the  PS fluid. Interestingly enough, however, this expectation only turns true for the highest repulsive barrier considered ($\epsilon^*=3$, where a  good agreement with MD data is observed in the density range $\phi<0.5$), as well as for $\epsilon^*=1$ and $\phi>0.8$.
Since, according to Eq.\ \eqref{2.13}, the Enskog-like values of $D^*$ are smaller than those obtained from the Boltzmann equation, the former values are worse than the Boltzmann values whenever the latter underestimate the MD data. As shown in Fig.\ \ref{fig1}, this is what happens for $\epsilon^*=0.2$ and $0.5$, as well as for $\epsilon^*=1$ and $\phi\lesssim 0.5$.

The system with $\epsilon^*=3.0$ deserves further comments. First, as indicated above, neither the Boltzmann equation nor the Enskog-like correction provides reliable values of $D^*$ for $\phi>0.6$. In fact, the MD values decay with increasing density much more rapidly than both theories predict. Moreover, the MD diffusion data (in the semi-logarithmic representation in Fig.\ \ref{fig1}) are seen to exhibit an inflection point near  $\phi=0.4$, a qualitative feature not accounted for by either the Boltzmann or the Enskog theories of the PS fluid. On the other hand, the existence of an inflection point of $\log D^*$ vs $\phi$ is also present in the HS system, as observed from the Enskog and the empirical lines in Fig.\ \ref{fig1}. In fact, the MD data of the PS fluid with $\epsilon^*=3$ are close to the HS values up to $\phi\approx 0.2$, analogous to what happens with the pressure (see Fig.\ \ref{Z}). For larger densities, however, the self-diffusion of  PS particles is much larger than that of HS particles at the same packing fraction.
One may suggest that, in this range of higher densities with a high repulsive interaction, the self-diffusion process is greatly affected by \emph{static} structural effects together with  \emph{dynamic} correlated motion involved in the PS model system. Later in this section, we will comment on some interesting observations from additional structural and dynamic results obtained from our MD calculations. It may be worthwhile noting here that, even in the simple HS system, there are various transition properties, mostly investigated by simulation approaches, such as  freezing of the fluid ($\phi =0.494$), melting of the solid ($\phi =0.545$), loose random packing ($\phi =0.555$), glass transition ($\phi =0.57$), and dense random packing ($\phi =0.64$) \cite{SH03}.

\subsection{Enskog $\chi$ factor and thermodynamic consistency}
\begin{figure}
\includegraphics[width=0.95\columnwidth]{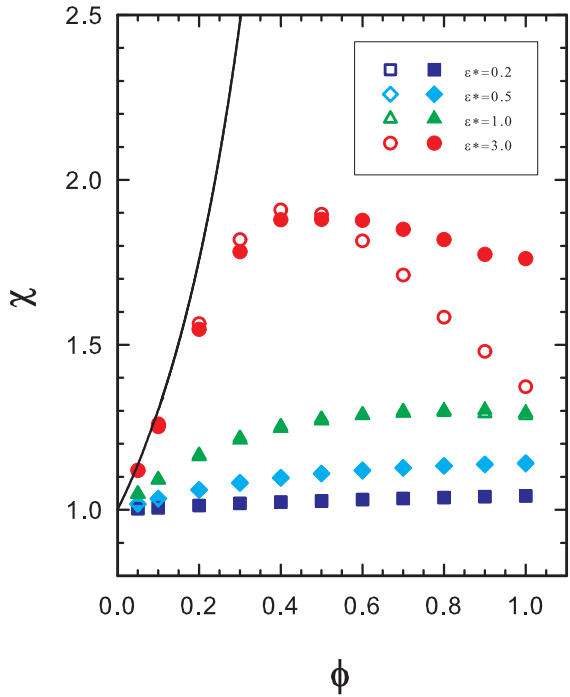}
\caption{(Color online) Enskog-correction factor $\chi$ as a function of packing fraction $\phi$ for several values of $\epsilon^*$, as obtained in MD simulations. The solid symbols are obtained from the contact-value method, while the open symbols are obtained from the compressibility factor. The solid line represents the function $\chi^\HS(\phi)$ given by the CS equation of state for the HS fluid [Eq.\ \protect\eqref{CS}].}
\label{fig2}
\end{figure}

In order to gain more detailed insights into the thermodynamic and structural properties related to the diffusion behavior, we have also examined other relevant properties of the PS model interaction system. Here, we start by considering the Enskog-type correction  $\chi$ factor.

According to statistical mechanics,  there are two main routes in theoretical and simulation studies to obtain $\chi$ in a PS system: (a) one  from the contact value of the radial distribution function as $\chi^\PS=g^\PS(\sigma^+)$ and (b) another one from the pressure via the so-called virial route as represented by Eq.\ \eqref{2.1}, i.e., $\chi^\PS=(Z^\PS-1)/4\phi(1-e^{-\epsilon^*})$. In simulation approaches, $\chi$ values in the contact-value method (a) can be directly computed during simulation runs, while  in the second method (b) they are  indirectly evaluated from  MC ensemble averages or  MD time averages of the compressibility factor at a given thermodynamic condition. In principle, except for unavoidable computational errors, those two methods should generate the same value  in \emph{equilibrium stable liquid states}.

As shown in Fig.\ \ref{fig2}, an excellent agreement between those two methods is observed over the entire density range for  $\epsilon^*=0.2$, $0.5$, and $1.0$.
Again, the system with $\epsilon^*=3.0$ requires separate comments. In that case, we observe that, up to $\phi\approx 0.2$  the MD  values are very close to the HS contact values obtained from the CS formula. This agrees with what is observed in Figs.\ \ref{Z} and \ref{fig1}. Nevertheless,    the HS values of $\chi$ increase rapidly as the density increases, while the PS  values reach a maximum around $\phi=0.4$ and decrease thereafter. The most striking feature observed in Fig.\ \ref{fig2} is the separation between the MD values of $\chi$ obtained from methods (a) and (b) with $\epsilon^*=3.0$ and $\phi>0.6$, with the maximum relative deviation being of almost $30$\% at $\phi=1$.

Also, although not shown in Fig.\ \ref{fig2},  we have evaluated the values of the ratio $g(\sigma^+)/[g(\sigma^-)e^{\epsilon^*}]$ by using the extrapolated MD data from $g(r)$. This ratio should take  the unity value, except for computational errors,  at any given density and repulsive energy parameter if the system is in an equilibrium stable liquid state. We have observed that the deviations  of $g(\sigma^+)/[g(\sigma^-)e^{\epsilon^*}]$ from unity in the cases  $\epsilon^*=0.2$ and  $\epsilon^*=0.5$  are less than $0.1$\% and $0.3$\%, respectively. The internal agreement between $g(\sigma^+)$ and $g(\sigma^-)e^{\epsilon^*}$ is also very good in the case $\epsilon^*=1.0$, with a maximum deviation of about $3$\% at $\phi=0.9$. Again, the least penetrable case ($\epsilon^*=3.0$) presents a peculiar behavior. Up to $\phi=0.5$ the ratio $g(\sigma^+)/[g(\sigma^-)e^{\epsilon^*}]$ deviates from $1$ less than $5\%$, but thereafter it markedly increases with density until having $g(\sigma^+)/[g(\sigma^-)e^{\epsilon^*}]\simeq 1.6$ at $\phi=1$.

All these discrepancies between the values of $\chi$ obtained from $g(\sigma^+)$, $(Z-1)/4\phi(1-e^{-\epsilon^*})$, and  $g(\sigma^-)e^{\epsilon^*}$ when $\epsilon^*=3.0$ and $\phi\gtrsim 0.6$  {are likely} due to the appearance of cluster-forming \emph{metastable} structures. Under these conditions we have employed the direct contact-value method $\chi=g(\sigma^+)$ to obtain the Enskog-type kinetic approximations for the self-diffusion coefficient  shown in Fig.\  \ref{fig1}.

\subsection{Effective particle volume fraction}
\begin{figure}
\includegraphics[width=0.95\columnwidth]{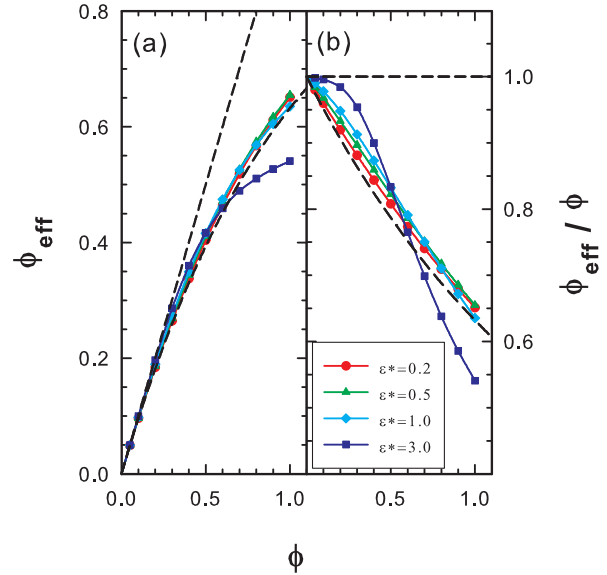}
\caption{(Color online) (a) Effective particle volume fraction $\phi_{\text{eff}}$ and (b) ratio $\phi_{\text{eff}}/\phi$ as  functions of the packing fraction $\phi$. The symbols (with lines to guide the eye) represent MD data for several values of $\epsilon^*$. The upper and  lower  {dashed} lines correspond to the limiting cases of the non-overlapping HS system ($\epsilon*\to\infty$) and the totally random overlapping PS system ($\epsilon*\to 0$), respectively.}
\label{fig3}
\end{figure}

As a dimensionless measure of the number of particles per unit volume we are using the nominal packing fraction $\phi=({\pi}/{6})n\sigma^3$. In the case of HS systems, $\phi$ coincides with the fraction of the total volume that is occupied by the spheres. {For PS systems, however,} particles can interpenetrate, thus reducing the fraction of occupied volume. Let us define the effective particle volume fraction $\phi_{\text{eff}}$ as the average effective total volume occupied by PS particles divided by the system volume. Of course, $\phi_{\text{eff}}$ is a function of both $\phi$ and $\epsilon^*$ that satisfies the inequality $\phi_{\text{eff}}\leq\phi$, with the equality taking place only in the HS limit ($\epsilon^*\to\infty$) or in the low-density limit ($\phi\to 0$).  For fully penetrable systems in the high-penetrability limit ($\epsilon^*\to 0$), the corresponding particle configuration will become that of a totally unbiased random structure, and this leads statistically to  $\phi_{\text{eff}}=1-e^{-\phi}$ \cite{SMM99}.

We have calculated $\phi_{\text{eff}}$ by the simple hit-and-miss method using a uniform {($10\sigma\times 10\sigma\times 10\sigma$)} grid over approximately half a million  equilibrium configurations during our MD computations. The results are displayed in Fig.\ \ref{fig3}. For the systems with $\epsilon^*=0.2$, $0.5$, and $1.0$, the resulting data are very close to (but slightly above) the curves representing randomly distributed configurations. In the case $\epsilon^*=3.0$, the MD values of $\phi_\text{eff}$ are close to $\phi$ up to $\phi\approx 0.2$, which indicates HS-like configurations in that density range. As the density increases further, $\phi_\text{eff}/\phi$ significantly decreases and at $\phi\approx 0.6$ it even crosses the random distribution expectation. It is paradoxical that at $\phi=1$, for instance, particles are so much overlapped in the case $\epsilon^*=3.0$ that less than $55$\% of the available volume is actually occupied by them; in contrast, at the same density, particles occupy about $65$\% of the total volume in the case of a much less repulsive barrier $\epsilon^*=0.2$. Again, this signals in the case $\epsilon^*=3.0$ a transition near $\phi=0.6$ to a metastable state characterized by a high degree of clustering.

\subsection{Mean free path {and collision frequencies}}
\begin{figure}
\includegraphics[width=0.95\columnwidth]{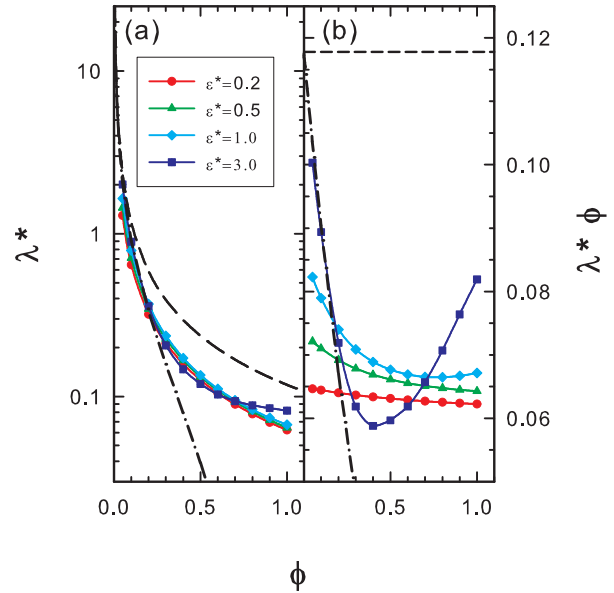}
\caption{(Color online) (a) Reduced mean free path $\lambda^*=\lambda/\sigma$ (in semi-logarithmic scale) and (b)  product $\lambda^*\phi$ as  functions of the packing fraction. The symbols (with lines to guide the eye) represent MD data for several values of $\epsilon^*$. The  {dashed} and dash-dotted lines correspond to the HS system in the Boltzmann and Enskog approximations, respectively.}
\label{fig4}
\end{figure}

In the HS kinetic theory, the mean free path $\lambda^\HS$ in the dense system is related to that in the high-dilution limit  $\lambda_0^\HS=\sigma/6\sqrt{2}\phi$ by the $\chi^\HS$ factor in a similar way as in Eq.\ \eqref{2.10}, i.e.,  $\lambda^\HS=\lambda_0^\HS/\chi^\HS$ \cite{CBDM77}. In this case of HS systems the mean free path simply characterizes the typical distance (much higher than the diameter $\sigma$ in the dilute regime) traversed by a sphere between two successive collisions. Of course, the distance between the centers of the two particles in a HS collision is $r=\sigma^+$. In contrast, two classes of events contribute to the mean free path in the PS system (see Fig.\ \ref{sketch}). On one hand, one still has hard collisions taking place at $r=\sigma^+$. On the other hand, soft encounters give rise to a primary ``external'' collision at $r=\sigma^+$ followed by a secondary ``internal'' collision at $r=\sigma^-$; this second event contributes to the mean free path with a value smaller than $\sigma$. Therefore, every soft collision has two contributions (primary and secondary) to the mean free path.
For dilute systems, where $\lambda\gg \sigma$, the primary contribution is on the order of $\lambda$, while the secondary one is practically zero{; as penetrability increases,} almost all the collisions are soft and thus the mean free path is about half the value obtained in a HS system at the same {(low)} density.

We have evaluated the mean free path for PS fluids during MD simulations, and the results are displayed in Fig.\ \ref{fig4}.  In the semi-logarithmic scale of Fig.\ \ref{fig4}(a), the  values of $\lambda^*\equiv \lambda/\sigma$  display a similar decaying behavior for the cases   $\epsilon^*=0.2$, $0.5$, and $1.0$. However, again in the case   $\epsilon^*=3.0$ a different behavior can be observed, where $\lambda^*$ is seen to exhibit a  weak density dependence for   $\phi>0.6$. Figure \ref{fig4}(b) shows that, in the case $\epsilon^*=3.0$, the product $\lambda^*\phi$ agrees with the HS prediction up to $\phi=0.2$, but then it presents an accentuated minimum at $\phi\simeq 0.4$; this peculiar behavior is another distinct evidence for cluster-forming structures. On the other hand, the curves for  $\epsilon^*=0.2$, $0.5$, and $1.0$ clearly depart from the HS values, even for small densities. This illustrates the penetrability effects in the PS fluid and the influence of soft collisions, as discussed above. For instance, in the case   $\epsilon^*=0.2$ the values of $\lambda^*\phi$   slowly decay with density, taking a nearly constant value that is almost half  the Boltzmann HS value $(\lambda_0^\HS/\sigma)\phi=1/6\sqrt{2}\simeq 0.11785$.

\begin{figure}
\includegraphics[width=0.95\columnwidth]{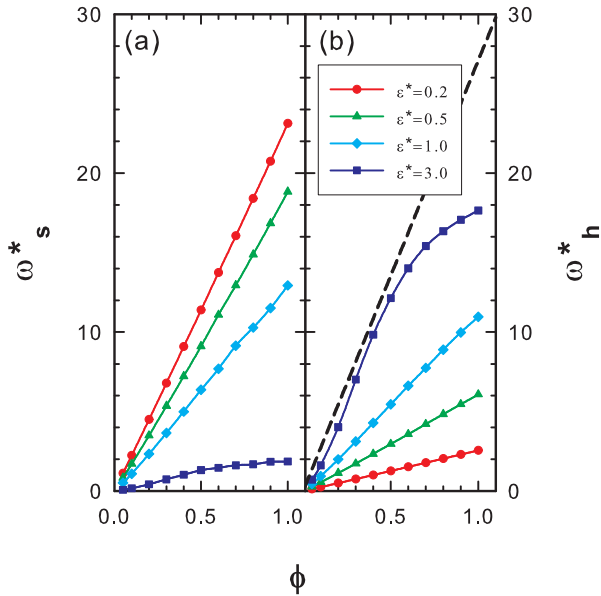}
\caption{(Color online) Reduced soft-penetrable collision frequency $\omega_s^*=\omega_s\sigma/\sqrt{k_BT/m}$ and (b)  reduced hard-reflective collision frequency $\omega_h^*=\omega_h\sigma/\sqrt{k_BT/m}$ as functions of the packing fraction $\phi$. The symbols (with lines to guide the eye) represent MD data for several values of $\epsilon^*$. The  {dashed} line in  (b) corresponds to the HS system, i.e., ${\omega^*}^\HS=(48/\sqrt{\pi})\phi$.}
\label{fig6}
\end{figure}

For HS systems, the collision frequency is $\omega^\HS=2\langle v\rangle/\lambda^\HS$, where $\langle v\rangle=\sqrt{8k_BT/\pi m}$ is the average speed. In the case of PS systems, it is instructive to decompose the collision frequency $\omega=\omega_s+\omega_h$ into the soft-type collision frequency ($\omega_s$) and the hard-type collision frequency ($\omega_h$).
{They are calculated as follows. Every time a collision with relative speed $v_{12}$ and impact parameter $b$ occurs, it is cataloged as either soft or hard depending on whether $(mv_{12}^2/\epsilon)[1-(b/\sigma)^2]-1$ is positive or negative, respectively. If $\mathcal{N}_s(t)$ and $\mathcal{N}_h(t)$ denote the total numbers of soft and hard collisions, respectively, over a time interval $t$, then $\omega_s=2\mathcal{N}_s(t)/t N$ and $\omega_h=2\mathcal{N}_h(t)/t N$, where $N$ is the total number of particles.}

{The collision frequencies $\omega_s$ and $\omega_h$} are plotted in Figs.\ \ref{fig6}(a) and \ref{fig6}(b), respectively.
We observe that  soft collisions are dominant  in the cases   $\epsilon^*=0.2$, $0.5$, and (to a lesser extent) $1.0$. In particular, soft collisions represent about $90$\% of all collisions for all the systems with $\epsilon^*=0.2$. It is interesting to note that both $\omega_s$ and $\omega_h$ grow almost linearly with $\phi$ in those three cases. In contrast, if $\epsilon^*=3.0$ we find an initial quasi-linear growth followed by a much slower regime for $\phi\gtrsim 0.6$. As might be expected, for $\epsilon^*=3.0$ hard collisions dominate over soft ones, with the former representing about $90$\% of all collisions.

\subsection{Velocity autocorrelation function {and mean square displacement}}
\begin{figure}
\includegraphics[width=0.95\columnwidth]{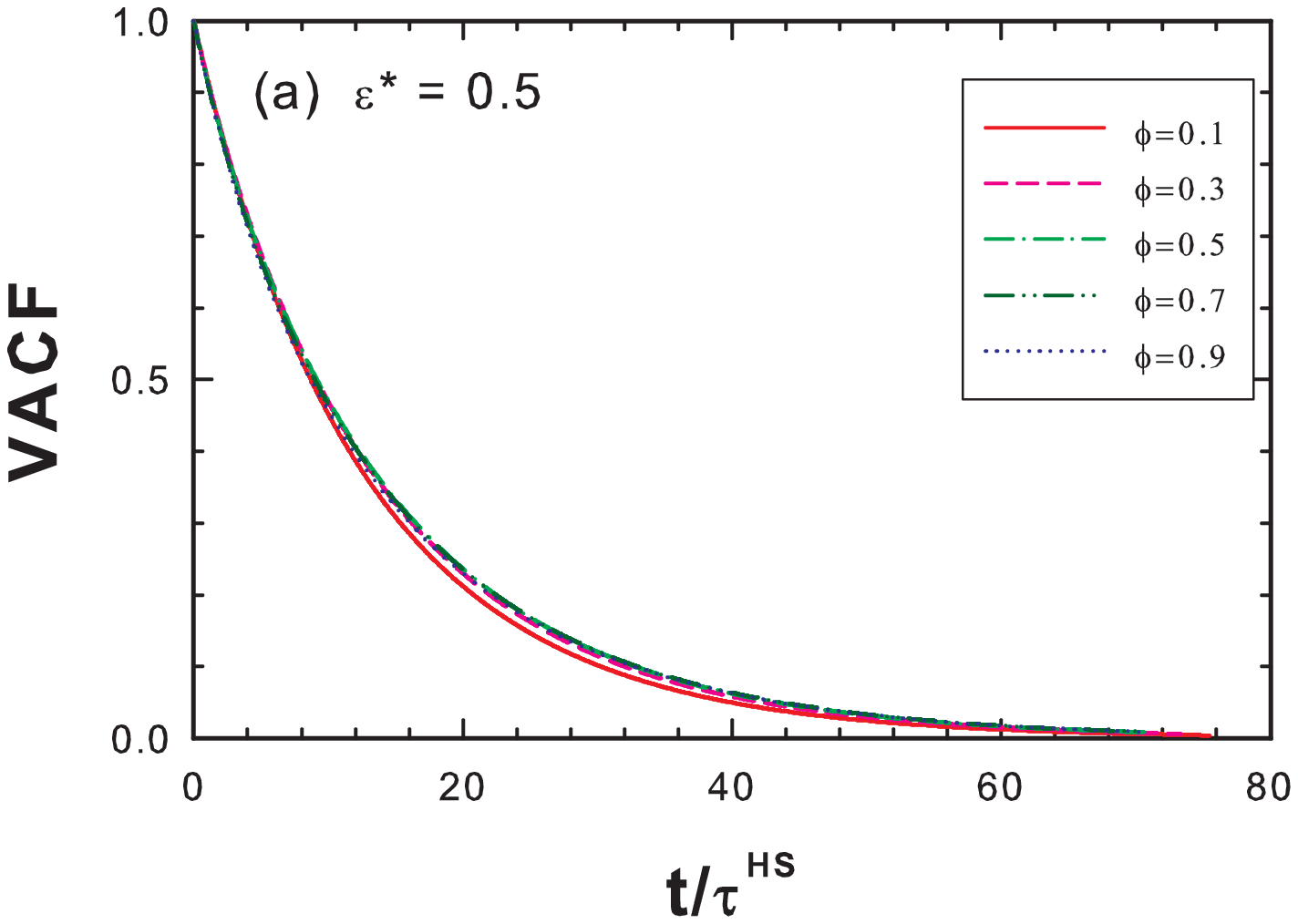}\\
\includegraphics[width=0.95\columnwidth]{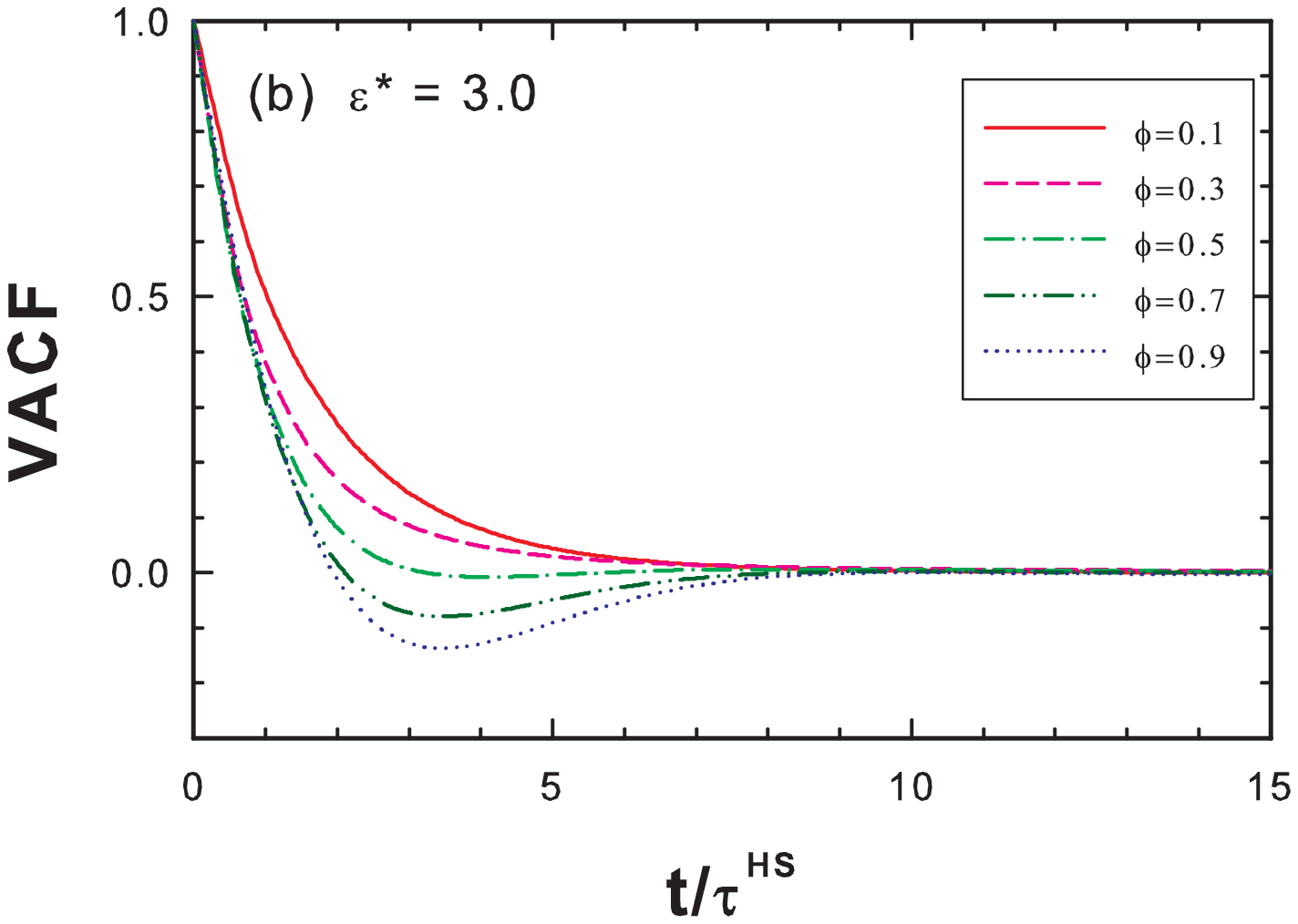}
\caption{(Color online) Normalized VACF as a function of the reduced time in units of $\tau^\HS$ for (a) $\epsilon^*=0.5$ and (b)  $\epsilon^*=3.0$. {The curves correspond to $\phi=0.1$ (solid lines), $0.3$ (dashed lines), $0.5$ (dash-dotted lines), $0.7$ (dash-dot-dot lines), and $0.9$ (dotted lines)}. Note the different horizontal scales in both panels.}
\label{fig7}
\end{figure}

Both the VACF [see Eq.\ \eqref{VACF}] and the MSD [see Eq.\ \eqref{MSD}] provide useful insights into the dynamic time-dependent behavior related to  diffusion processes  in PS systems.
It is then illustrative to analyze the manner in which those functions change with increasing densities and repulsive energy parameters.

In Figs.\ \ref{fig7}(a) and  \ref{fig7}(b) we display the VACF (in units of $3k_BT/m$) as a function of time in units of the relevant HS mean collision time $\tau^\HS=1/\omega^\HS=\lambda^\HS/2\langle v\rangle$  for $\epsilon^*=0.5$ and $\epsilon^*=3.0$, respectively, and several characteristic densities.
The primary mechanism involved in the rapid decay of the VACF is provided by hard collisions, so that colliding particles rapidly forget their initial velocities through successive collisions. For soft-penetrable collisions, post-collision velocities (or, equivalently, the colliding particle trajectories) are relatively correlated with their own initial values. In agreement with this, the normalized VACFs for the systems with  $\epsilon^*=0.5$ exhibit  similar exponentially decaying behaviors for different densities, since the soft-penetrable collision process is dominant in this case. As a consequence the areas below the curves are hardly dependent on $\phi$, which implies $D\sim \tau^\HS\propto \phi^{-1}$.
In the case $\epsilon^*=3.0$ the resulting VACF for  $\phi=0.1$ exhibits a decaying exponential behavior similar to that of  the system with $\epsilon^*=0.5$ and the same density, except that now the decay is much more rapid due to the prevalence of hard collisions. For higher densities, the development of  cluster-forming structure  significantly influences the collective motion of penetrable particles by  retardation (for incoming  particles) or acceleration (for outgoing particles) of the velocities of colliding pairs. Under those conditions the resulting VACFs exhibit a non-exponential behavior. Furthermore, the trajectories of colliding particles are largely restricted by backscattering or cage effects between clusters in the higher-density regime ($\phi>0.6$ and  $\epsilon^*=3.0$), as shown in Fig.\ \ref{fig7}(b).

\begin{figure}
\includegraphics[width=0.95\columnwidth]{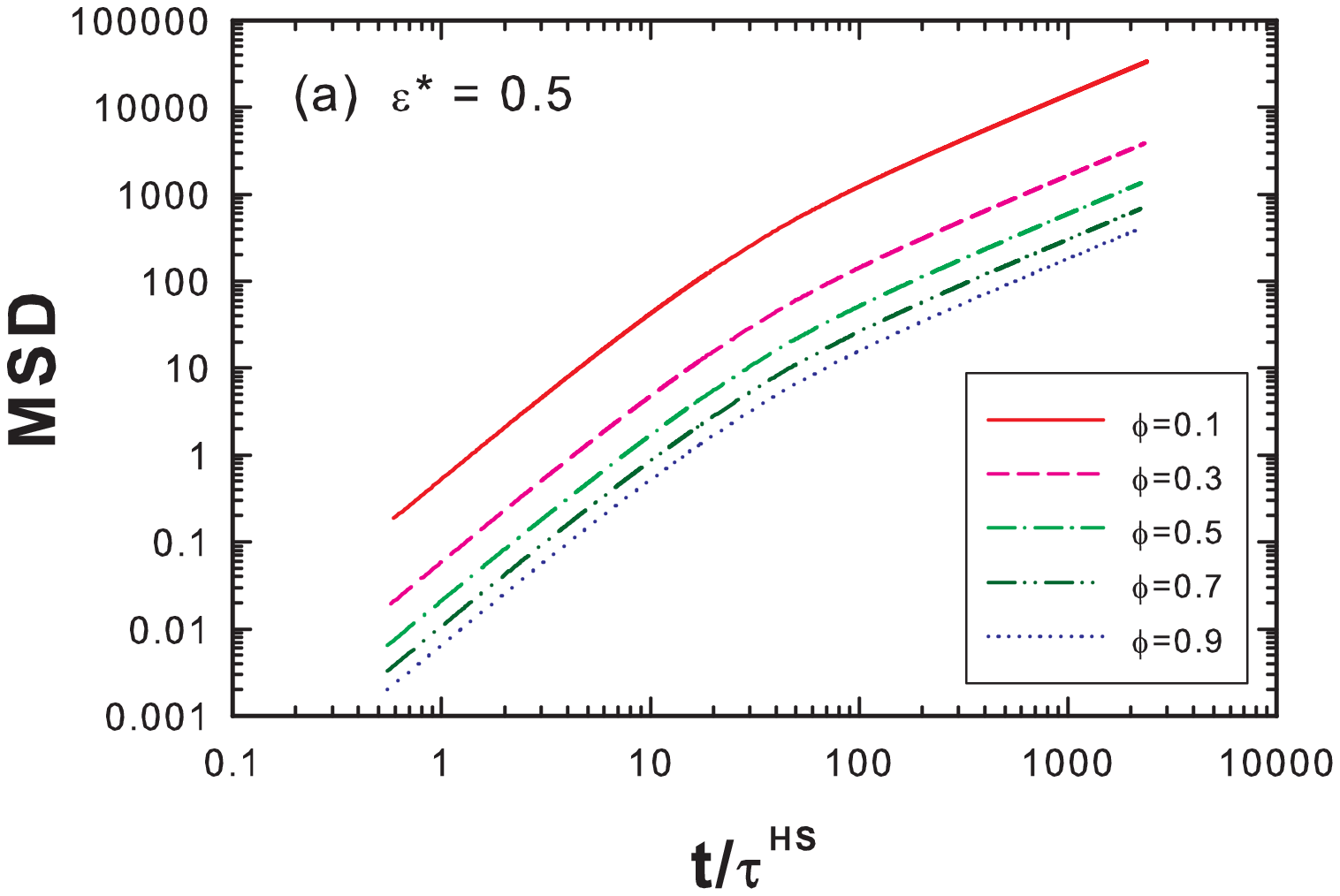}\\
\includegraphics[width=0.95\columnwidth]{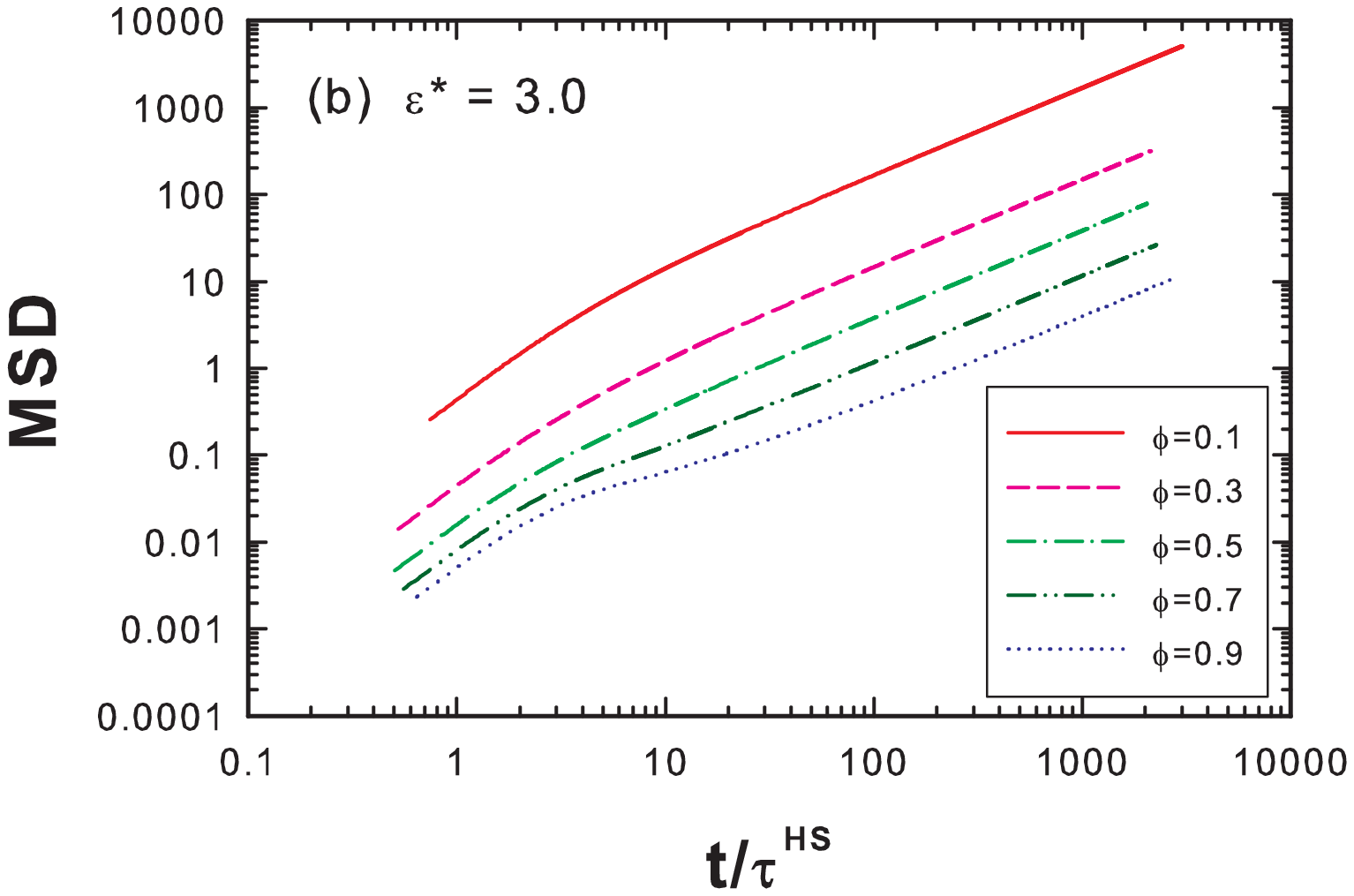}
\caption{(Color online) Log-log plot of the MSD (in units of $\sigma^2$) as a function of the reduced time in units of $\tau^\HS$ for (a) $\epsilon^*=0.5$ and (b)  $\epsilon^*=3.0$. {The curves correspond to $\phi=0.1$ (solid lines), $0.3$ (dashed lines), $0.5$ (dash-dotted lines), $0.7$ (dash-dot-dot lines), and $0.9$ (dotted lines)}.}
\label{fig8}
\end{figure}

As displayed in the log-log plot of Fig.\ \ref{fig8},  trends similar to the ones  mentioned above for the VACF can be observed from the MD results for the MSD curves. At very short times, before hardly occurring particle collision, the  MSD curves changes more rapidly than in longer times. This is due to the ballistic motion of  particles until they collide with their neighbors, resulting in $\langle r^2(t)\rangle\sim t^2$. Subsequent collisions make  trajectories resemble a random-walk diffusion process, so that, after a certain transition period, the diffusive regime defined by Eq.\ \eqref{MSD} is established. These behaviors are clearly illustrated in Figs.\ \ref{fig8}(a) and \ref{fig8}(b), except that the  MSD curve for  $\phi=0.9$ and  $\epsilon^*=3.0$ presents a \emph{transient}  anomalous \emph{sub-diffusive} behavior where $\langle r^2(t)\rangle\sim t^\gamma$, with $\gamma<1$. This signals a hindrance of the diffusion process by obstruction or trapping phenomena. Once the diffusive regime ($\gamma=1$) is reached  for longer times, it is characterized by a very low value of the self-diffusion coefficient, as shown in Fig.\ \ref{fig1}.

\subsection{Discussion on the Boltzmann kinetic theory}
Before concluding this section, it is of interest to return to one of the observations made earlier to explain some relevant shortcomings involved in the Boltzmann theoretical approximation. As illustrated in Fig.\ \ref{fig1}, together with Figs.\ \ref{fig7} and \ref{fig8}, the failure of the Boltzmann kinetic approximation for the PS model fluid becomes more important as the density increases. This is not surprising: one may recall that the Boltzmann kinetic theory is based on the high-dilution limit. A key element related to this kinetic theory is the molecular chaos assumption, known as ``Stosszahlansatz,'' in which the pre-collision velocities of two colliding particles are assumed to be totally uncorrelated. In addition, regardless of a given model potential, the Boltzmann kinetic theory deals with only binary collision effects by totally neglecting multiple collisions. As observed in MD simulations for the PS model potential in this work, the deviation between our MD diffusion data and the Boltzmann predictions can be largely due to the neglect of such spatio-temporal correlations in the PS collision dynamics, particularly in dense system with cluster-forming structures. For example, a simple conjectural argument will intuitively give, in analogy with the HS relation $D^\HS\propto\lambda^\HS$, that particles with larger mean free paths become more diffusively dispersed (larger diffusion coefficients), and vice versa. However, for the dense PS fluid with the highest repulsive barrier ($\phi >0.6$ and  $\epsilon^*=3.0$), our MD results clearly manifest contradictions against this simple conjecture:  similar  values of the mean free path (Fig.\ \ref{fig4}) yield a significant reduction in the corresponding self-diffusion coefficient (see Fig.\ \ref{fig1}). This  partly explains that the Boltzmann kinetic theory does not take proper account of the consequences of correlated collisions on the self-diffusion coefficient in the PS  system, as investigated in this work. More detailed MD simulation studies can be very helpful to enable qualitative predictions of the underlying behavior of the PS interaction systems together with statistical-mechanical approaches, which will be one of our current research topics for further theoretical and simulation work.

\section{Concluding remarks \label{sec5}}

In this work, as an intermediate between theory and experiment, MD simulations have been carried out to investigate the detailed diffusion behavior of PS model fluids. The self-diffusion coefficient has been calculated from its related time-dependent properties of the VACF and the MSD. The resulting simulation data have been used to assess theoretical predictions by the Boltzmann kinetic equation and an Enskog-like correction. Detailed insights involved in the cluster formation for penetrable spheres have been  observed from the effective particle volume fraction, the mean free path, and the collision frequency for both the soft-type penetrable and the hard-type reflective collisions.

{A reasonable good agreement with Boltzmann and Enskog theoretical approximations is found in the cases   $\epsilon^*=0.2$, $0.5$, and $1.0$. On the other hand, for the dense PS fluid with the highest repulsive barrier ($\phi >0.6$ and  $\epsilon^*=3.0$), several distinct evidences of the cluster-forming structure are exhibited from static structural and dynamic collisional properties. In that case, a poor agreement between theory and simulation is observed due to those effects, especially correlated collision processes.} Under those conditions, the indirect virial route and the direct contact-value route to obtain the Enskog-type correction factors yield inconsistent results. This indicates that for  $\epsilon^*=3.0$ and $\phi >0.6$ the states reached in our MD simulations are not of strict thermodynamic equilibrium but are long-lived metastable states. We are currently examining such cluster-formation conditions with larger numbers of penetrable spheres to check the number dependence of transport properties including the self-diffusion, the shear viscosity, and the thermal conductivity coefficients.

\begin{acknowledgments}
S.-H.S. wishes to express his thanks to  Jae-Moon Yang,  Young-Jin Ha, and  Jong-Hoon Sohn for their help and implementations during simulation runs.
The research of A.S. was supported by the Ministerio de Educaci\'on y Ciencia (Spain) through Grant No.\
FIS2007-60977 (partially financed by FEDER funds) and by the Junta
de Extremadura through Grant No.\ GRU10158.
\end{acknowledgments}

\end{document}